\documentclass[runningheads]{llncs}
\usepackage{graphicx}
\usepackage{cite}
\begin{document}
\title{Interpersonal distance in VR: reactions of older adults to the presence of a virtual agent}
\titlerunning{Older adults' responses to the virtual agent}
%
\author{Grzegorz Pochwatko\inst{1}\orcidID{0000-0001-8548-6916}\and
Barbara~Karpowicz\inst{2}\orcidID{0000-0002-7478-7374
}\and
Anna~Chrzanowska\inst{1}\orcidID{0000-0002-2412-4169}\and
Wiesław~Kopeć\inst{2}\orcidID{0000-0001-9132-4171} }
\authorrunning{G. Pochwatko et al.}
%
\institute{Institute of Psychology, Polish Academy of Sciences
\email{gp@psych.pan.pl}\\
\url{https://psych.pan.pl} 
\and
Polish-Japanese Academy of Information Technology
\email{kopec@pja.edu.pl}\\
}
\maketitle              
\begin{abstract}
The rapid development of virtual reality technology has increased its availability and, consequently, increased the number of its possible applications. The interest in the new medium has grown due to the entertainment industry (games,  VR experiences and movies). The number of freely available training and therapeutic applications is also increasing. Contrary to popular opinion, new technologies are also adopted by older adults. Creating virtual environments tailored to the needs and capabilities of older adults requires intense research on the behaviour of these participants in the most common situations, towards commonly used elements of the virtual environment, in typical sceneries. Comfortable immersion in a virtual environment is key to achieving the impression of presence. Presence is, in turn, necessary to obtain appropriate training, persuasive and therapeutic effects. A virtual agent (a humanoid representation of an algorithm or artificial intelligence) is often an element of the virtual environment interface. Maintaining an appropriate distance to the agent is, therefore, a key parameter for the creator of the VR experience. Older (65+) participants maintain greater distance towards an agent (a young white male) than younger ones (25-35). It may be caused by differences in the level of arousal, but also cultural norms. As a consequence, VR developers are advised to use algorithms that maintain the agent at the appropriate distance, depending on the user's age.

\keywords{virtual reality  \and embodied virtual agents \and older adults.}
\end{abstract}
\section{Theoretical context}
\subsection{Serious virtual environments}
Virtual environments are a new communication medium. Interactions in VR have become particularly common in recent years when the quality of sound and graphics offered by modern devices has come close to natural (see \cite{bailenson2003interpersonal}, \cite{defanti2000better}). Nowadays, they are not just an amusing novelty. There are more and more serious applications and virtual experiences emerging. Some of them rely on interpersonal mediated communication, but, thanks to advanced algorithms, natural language processing and machine learning, an increasing number of apps incorporate embodied virtual agents. Users got used to intelligent virtual agents. Google Assistant or Siri is becoming a part of our everyday life, and the popularity of Alexa and Google Home devices is increasing (to name just a few most popular ones). But embodied agents are something more than that. People have a tendency to treat computers as if they were other people. Reeves and Nass \cite{reevesNass1996} show that users treat computers in a social and natural way. Giving them a body that responds naturally to the environment by mimicking human-like movement, reacting to the user's presence, following with the gaze and maintaining social distance, makes this tendency even stronger. One doesn’t have to imagine the ghost in the machine any more and suspend his/her disbelief. It’s (almost) not an illusion of copresence any more as well. Instead, we have (almost) genuine social interaction. One of the recent applications is shown by Kim et al. \cite{kim2018does} who proposed giving intelligent assistants a human body with the use of augmented reality. According to the expectations, contact with embodied agents resulted in higher confidence, trust and feeling of social presence.

Serious games, training and therapy applications are increasingly used in place of traditional methods. They are cheaper, sometimes more effective, they offer users almost unlimited access, especially where it is difficult for any reason. It is particularly important in times of pandemic when traditional services are limited and safer, remote methods are preferred. This refers primarily to older adults, who are highly vulnerable. However, serious games must constantly evolve to adapt to the changing conditions and needs of new user groups \cite{fleming2017serious}.

\subsection{Older adults' population} Older adults are the group in which the increase in interest in VR and experience with the use of various applications is the largest (according to the Global Web Index from 0.5\% in 2016, an increase to 6\% in 2018). The global virtual reality software market is predicted to double in size till 2023. Simultaneously, the percentage of older adults in the European population (according to the United Nations Population Division) will exceed 25\%. Similarly, the World Health Organization is predicting the dawn of the super-aged society. We will face the possible decline in both physical and mental health, which may lead to the deterioration of quality of life and well being of older adults. Can VR reduce those negative effects and assure active ageing?

At the same time, older adults can gain a lot from using applications developed in new technologies. This is not only about entertainment, which has dominated recently, but also about improving mental well-being and general functioning (e.g. \cite{garcia2015using}, \cite{korn2018using}, \cite{amorim2018virtual}). It seems we have a pretty good foundation for this. According to Syed-Abdul et al. \cite{syed2019virtual}, older adults have positive attitudes towards the use of VR as a supportive technology. Perceived ease of use is high, which makes the experience enjoyable and regraded as useful. These are the main factors influencing the intention to use. Complying to social norms, on the other hand, is less important.

\subsubsection{Objective} One of the critical challenges is to develop guidelines for creating effective interfaces in virtual environments targeted at older adults. An element of the VR interface can be a virtual agent, i.e. a humanoid representation of an application or artificial intelligence algorithms. An important task is to find out if there are any differences between younger and older adults in interacting with the virtual agency. An important parameter is a distance to the virtual agent. Maintaining a comfortable distance to the user may translate into the effectiveness of a training or therapeutic application. In the world of physical reality, interpersonal distance is an essential feature of individuals' social behaviour in relation to their physical environment and social interactions (for a review see \cite{sorokowska2017preferred}). The same is true of virtual reality \cite{bailenson2003interpersonal}.

\section{Method}
\subsection{Participants} Participants were invited to take part in the study conducted in the Virtual Reality and Psychophysiology Lab of the Institute of Psychology, Polish Academy of Sciences. Older adults were recruited from among the participants of VR workshops organised by the KOBO Association with XR Lab, Polish-Japanese Academy of Information Technology (PJAIT) and Emotion Cognition Lab of SWPS University. The VR workshops were organised as a part of follow-up studies of previous research on VR experience with older adults by Living Lab Kobo and HASE research initiative (Human Aspects in Science and Engineering) \cite{kopec2019vr}, based on previous intergenerational participatory studies by Living Lab PJAIT \cite{kopec2017living, kopec2017older}. Six older adults, aged 65+, were in the experimental group. There were three women and three men among them. The control group consisted of 6 people aged 18-35 (4 women and 2 men), volunteers from VRLab IP PAN participants' panel. The results of 12 people were analysed. All participants had a normal or corrected vision. All participants in the experimental group experienced immersive VR for the first time. The participants in the control group were selected to meet the same condition. People in the experimental and control groups were recruited in such a way that they differed only in belonging to a specific age group. The reason for this was that according to Sorokowska et al. \cite{sorokowska2017preferred}, gender and age are the only personal characteristics associated with interpersonal distance. The third feature, closing the list of significant predictors, is the air temperature (which in our case, in laboratory conditions, was kept at a constant level of 22 degrees Celsius). 

\subsection{Apparatus} Lenovo ThinkSTation P510 workstation with Intel (R) Xeon (R) E5-1630 3.70 GHz processor with 32 GB RAM, with Nvidia GeForce GTX 1080 Ti graphics card, and 64-bit Windows 10 Pro operating system was used. The virtual environment was prepared in the Worldviz Vizard Virtual Reality software, displayed using HMD HTC Vive with AudioStrap (AMOLED screen, 2160x1200 resolution, i.e. 1080x1200 per eye, 90Hz refresh rate, 110 degrees field of view width). It is worth noting that using a worse or much better display could affect the results.

\subsection{Procedure} The participants performed initial activities aimed at getting used to the virtual environment and the way of moving around. Display parameters, IPD, sharpness, HMD mount and controllers were also adjusted.

\begin{figure}
\includegraphics[width=\textwidth]{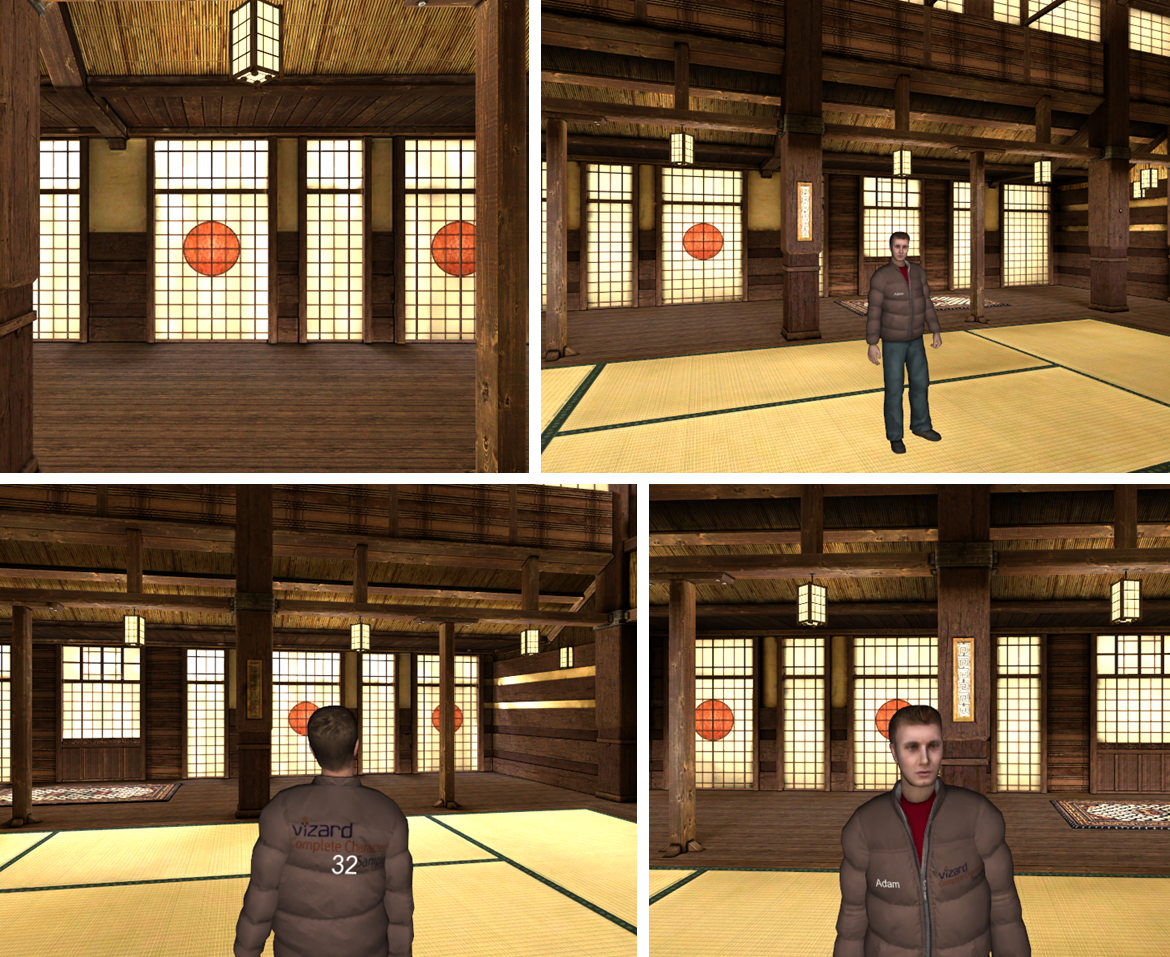}
\caption{The view of the immersive virtual environment that participants saw (top left: initial view, introduction and practice; top right: the beginning of the procedure; bottom left and right: close-ups of an agent.} \label{fig1}
\end{figure}

The participants were immersed in a virtual environment which was a mapping of a room corresponding to the dimensions of the laboratory's physical space. We used the standard HTC VIVE play area: 3.5 by 3.5 meters. The physical space was slightly larger to ensure the safety of the participants. On each side of the play area, we have left at least 1.5 meters of space to the nearest barrier. One virtual agent, a white young man, was placed in the room. The agent did not interact with the participants. We used the so-called idle pose animation, in which the character stands still and performs only small natural movements. More specifically, it is a cyclical animation in which the character stands relaxed, you can see breathing movements, slight movements of the head, arms, hands along the body, the center of gravity is shifted to the left or right side. The target task of the participants was to approach the agent and read the inscriptions on badges placed on his clothes, on the front and the back (cf. Bailenson et al. \cite{bailenson2003interpersonal}). The lettering on the agent's chest uses a font 3 cm high so that it cannot be read from the starting position. The inscription on the back uses an 8 cm high font. In both cases, white letters with black borders were used. The position of the participants' heads and hands was measured during the task (with a frequency of 90 Hz). It should be noted that traditional research on social distancing differs in approaches and methods from research using immersive VR. In the traditional study cited above, conducted in 42 countries, participants marked the declared distance towards a stranger, acquaintance and a close person on a paper diagram showing schematic drawings of two human figures, representing the participant and another person. This approach is convenient for mass intercultural research. In a virtual environment, the participant (in the case of this study) or the agent physically moves through space and adopts convenient positions. In our opinion, this type of measurement is more reliable and accurate. It’s worth noting that there are studies on social distance, also carried out in immersive VR, that contain elements such as the movement of an agent in the space of a motionless participant or instructions for the participant to take specific positions in space of an agent. Undoubtedly they may cause discomfort to the participants, which can be registered objectively, in the form of physiological response (HR and skin conductance changes resulting from activation of the sympathetic nervous system) or declaratively. For ethical reasons, we consciously resigned from procedures that could cause discomfort to participants.

\section{Results}
The research was preliminary, and therefore a small group of volunteers took part. In this situation, it can be assumed that the distribution of results will differ from normal. The Shapiro-Wilk distribution tests confirmed it for the group of seniors (for body distance: W\textsubscript{older adults} = .76, p$<$.05; W\textsubscript{control} = .91, p = .41; for hand distance: W\textsubscript{older adults} = .90, p = .36; W\textsubscript{control} = .86, p = .18). Failure to meet the assumption of normal distribution forces the application of the non-parametric Wilcoxon test for two independent samples.

\begin{figure}
\includegraphics[width=\textwidth]{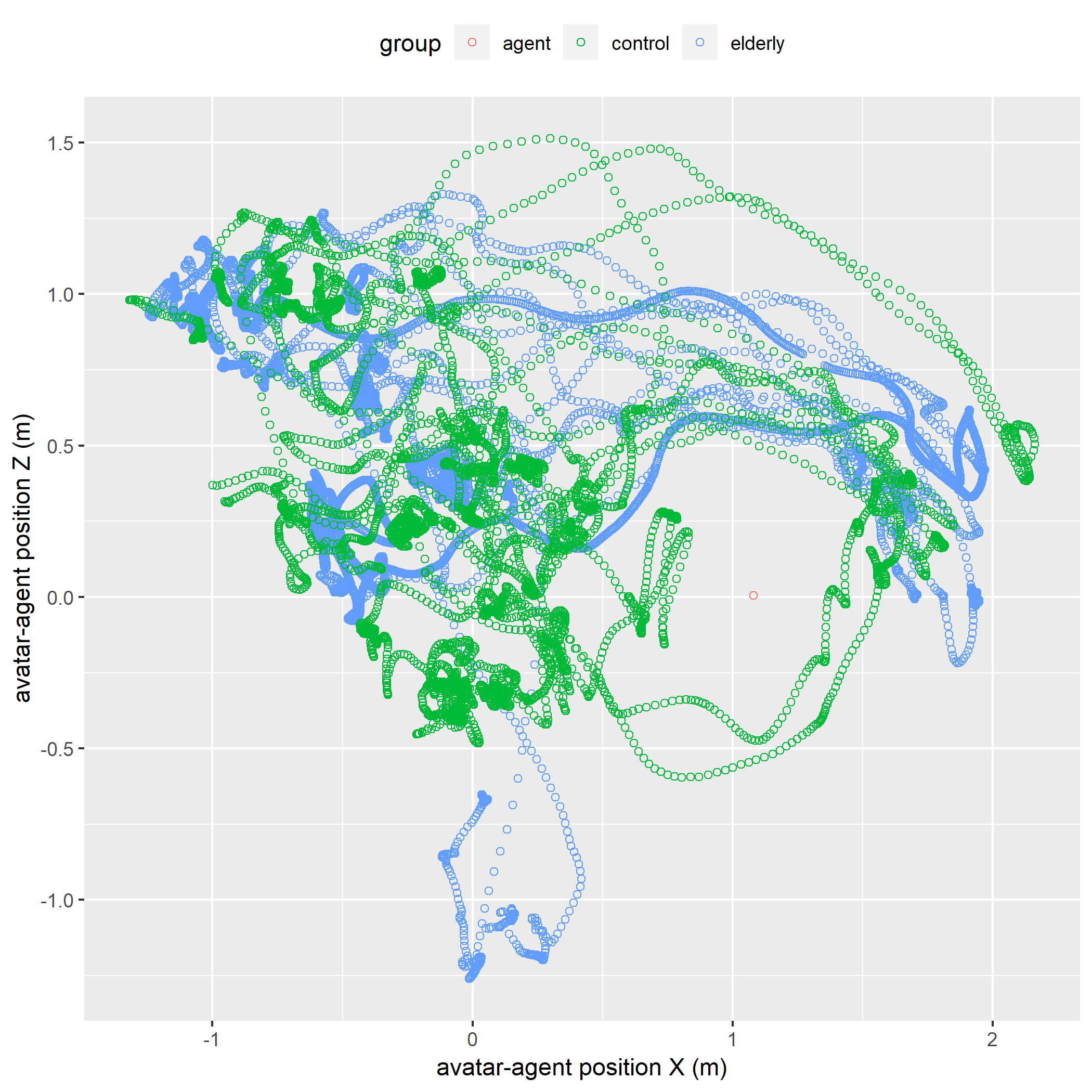}
\caption{Position towards embodied virtual agent (m).} \label{fig2}
\end{figure}

At first glance, the manner in which the older participants and the control group move in the space differs. The older adults move more slowly and cautiously than younger people from the control group. Their relative position towards the embodied virtual agent also differs (see Fig.~\ref{fig2}).

\begin{figure}
\includegraphics[width=\textwidth]{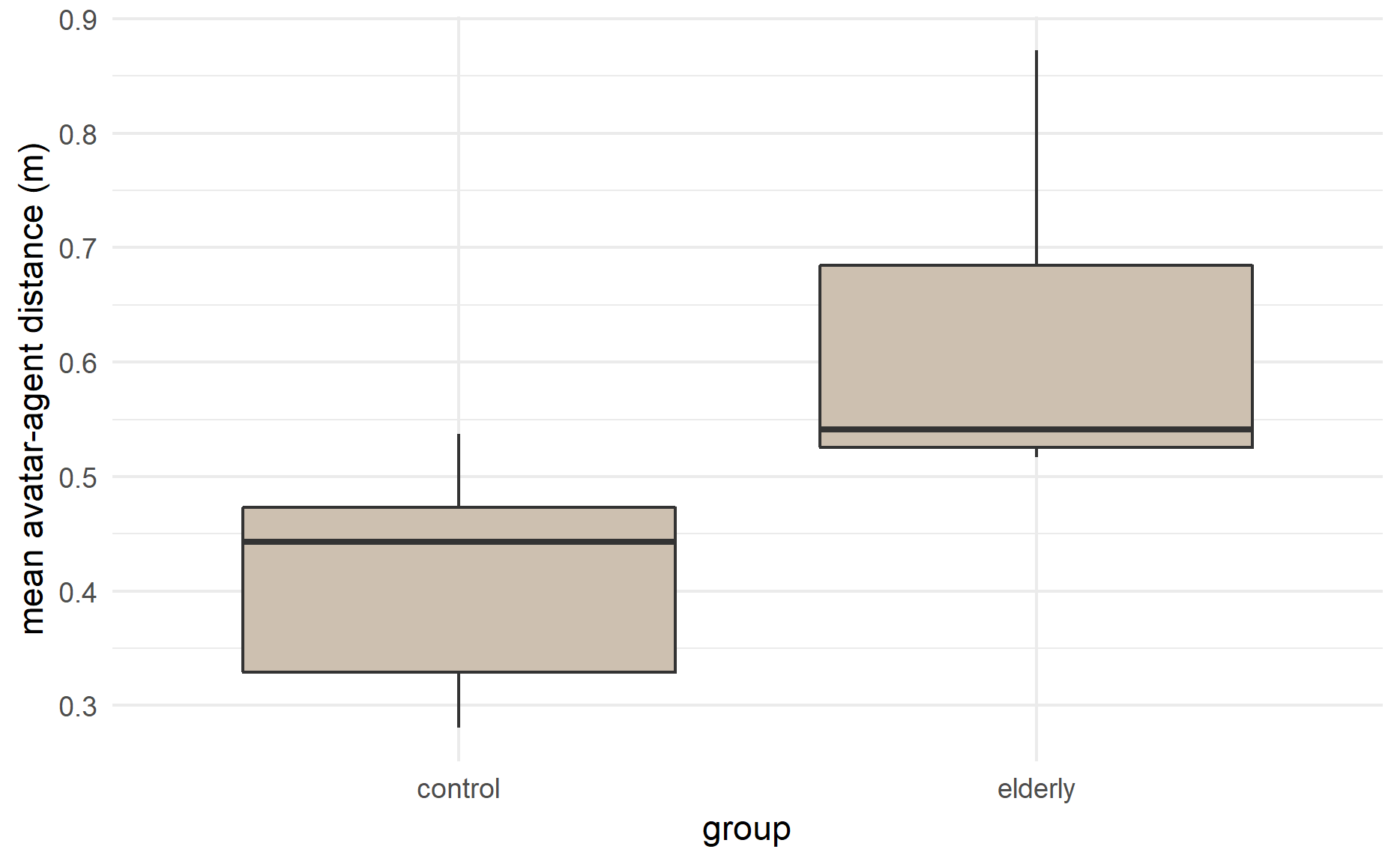}
\caption{Mean minimal body distance (m) towards embodied virtual agent in older adults and control group.} \label{fig3}
\end{figure}

The Wilcoxon test for independent samples revealed that older adults keep a greater distance from the virtual agent than younger people in the control group (0.54 m vs 0.38 m). The Wilcoxon test for two independent samples was found to be significant (W = 2, p $<$ .05; see Fig.~\ref{fig3}). 

\begin{figure}
\includegraphics[width=\textwidth]{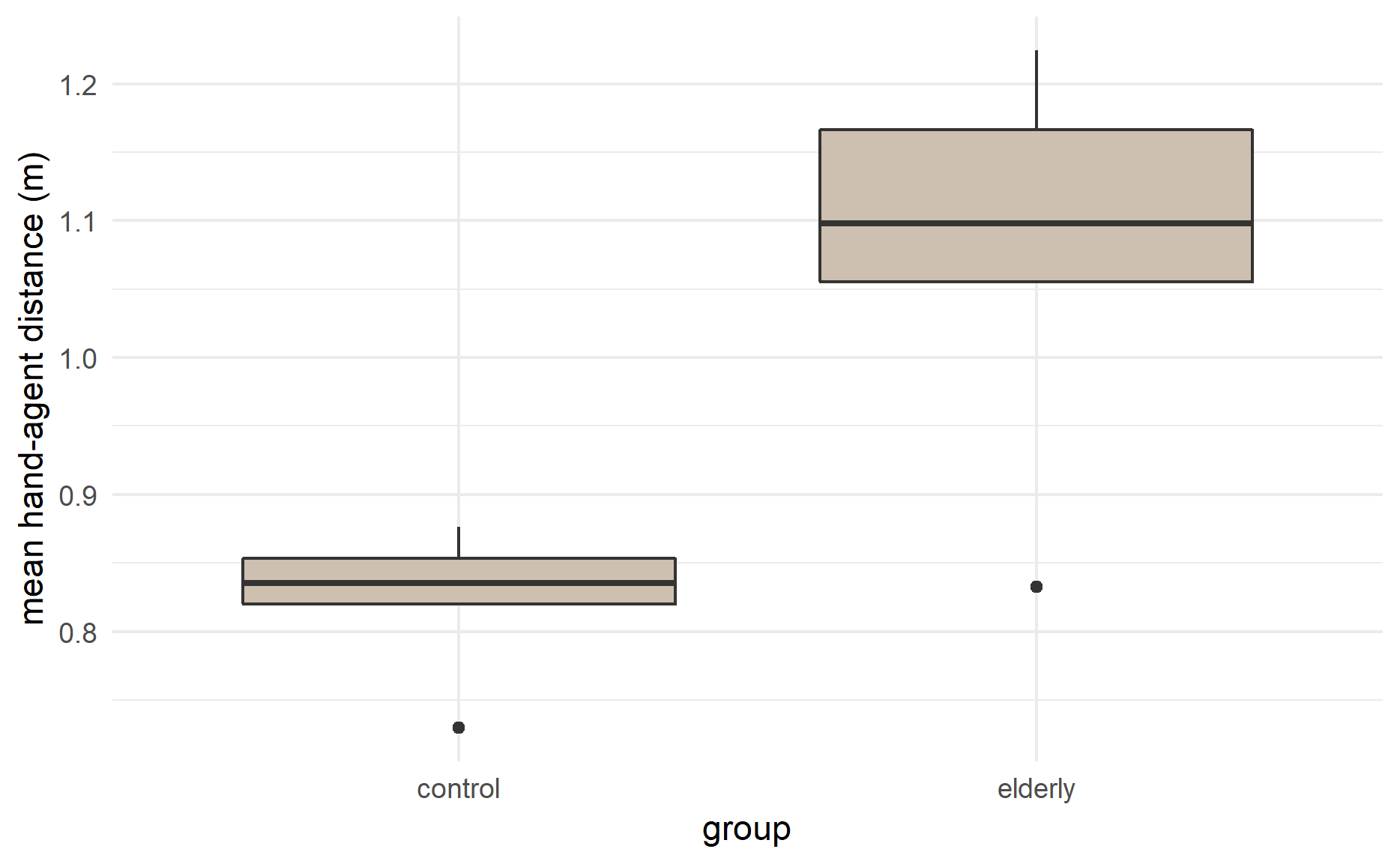}
\caption{Mean minimal hand distance (m) towards embodied virtual agent in older adults and control group.} \label{fig4}
\end{figure}

The same relationship occurred in the case of the distance between the hands and the embodied virtual agent. The Wilcoxon test for independent samples revealed that older adults keep the hands further from the virtual agent than younger people in the control group. The Wilcoxon test for two samples was found to be significant (W = 3, p$<$.05; see Fig.~\ref{fig4}). An analysis of the position of the hands in relation to the body showed that the participants did not try to stretch them towards the virtual agent. There were also no attempts to touch the agent's virtual body in any of the groups.

\section{Discussion}
Differences in the pace and manner of movement of the older and younger participants can be explained in two ways: first, in the case of older adults, there may be deficits in mobility and the sense of balance. Secondly, the lower speed and noticeable caution of movements could also result from the general less familiarity of the older participants with virtual environments. However, it should be noted that both the older adults and those in the control group experienced virtual reality for the first time in their lives.
It was possible to repeat and extend the results of Bailenson et al.\cite{bailenson2003interpersonal}. As for the interpersonal distance maintained, the participants treated the virtual agent in the same way that strangers are usually treated in a public place. Older participants avoided entering the agent's intimate space ($<$45 cm) and kept the average distance above 50 cm. The younger participants maintained a significantly smaller distance; they slightly entered the agent's intimate space (38 cm on average).
Hand position analysis showed that participants from both groups did not try to touch the agent. This may indicate that the virtual agent is treated like another living person in the same room. However, an alternative explanation cannot be ruled out. In case of negative emotions towards the virtual agent, it could be an aversive reaction.
Differences between younger and older participants in the distance kept from the virtual agent may have a cultural background \cite{sorokowska2017preferred}. Another possible explanation is that younger people have greater experience with virtual environments (such as video games) and virtual agents (such as non-player characters - NPCs), which may lead to them to treat the agent in the simulation in a more objective manner. Maybe the older adults are more under the illusion of social presence (or copresence). However, this requires more systematic research and larger trials.

\section{Implications and further directions}
There are many VR applications on the market that are designed for older adults (although most often not only about them). Among them, there are relaxation applications that allow one to regain peace and move into a pleasant, relaxing space, even when it is impossible in the real world. There are also virtual social environments that allow you to stay in touch with family and friends, and spend time together. A special place among them is occupied by applications aimed at residents of retirement homes or other such communities, thanks to which members can jointly participate in specially prepared simulations. Thanks to some, you can explore the world and watch various exciting places both in the form of spherical photos and 360 films, and even travel back in time and explore the same places at different times in history. The increasing percentage of older adults in the population, as well as their noticeably greater activity, will raise the interest in this type of solutions. The creators of VR applications should take into account the preferences of users if they want to provide them with a comfortable experience of their simulations. In applications whose interface is an embodied virtual agent, care should be taken to increase the distance towards the older user, as well as to avoid gestures that violate the intimate zone and virtual touch, if it is not accepted in a given culture. For younger users, it is possible to shorten the distance.
The use of VR applications by older adults requires further in-depth research on much larger trials. It is worth taking a closer look at the issues of reaction to facial expressions, and pantomimics of a virtual agent, gaze tracking, voice communication, issuing and listening to commands, compliance without pressure. An interesting development of the described research is also the manipulation of gender and age of a virtual agent.
\section{Acknowledgments}
We would like to thank our participants, older adults from our Living Lab, those affiliated with Kobo Association and volunteers from VRLab IP PAN panel who participated in this study. We would also like to thank all transdisciplinary experts involved with the HASE research initiative (Human Aspects in Science and Engineering).

\bibliographystyle{splncs04}
\bibliography{bibliography}
\end{document}